\documentclass[conference]{IEEEtran}
\IEEEoverridecommandlockouts
\usepackage{cite}
\usepackage{amsmath,amssymb,amsfonts}
\usepackage{algorithmic}
\usepackage{graphicx}
\usepackage{textcomp}
\usepackage{xcolor}
\usepackage{xspace}
\usepackage{macros}
\def\BibTeX{{\rm B\kern-.05em{\sc i\kern-.025em b}\kern-.08em
    T\kern-.1667em\lower.7ex\hbox{E}\kern-.125emX}}
\begin{document}

\title{Comparing Temporal Deep Learning Architectures for Discrete Cross-User Myoelectric Control}
\author{
    \IEEEauthorblockN{Ethan Eddy\IEEEauthorrefmark{1},
                      Evan Campbell\IEEEauthorrefmark{1},
                      Scott Bateman\IEEEauthorrefmark{2},
                      Erik Scheme\IEEEauthorrefmark{1}}
    
    \IEEEauthorblockA{\IEEEauthorrefmark{1}Department of Electrical Engineering, University of New Brunswick, Canada}
    \IEEEauthorblockA{\IEEEauthorrefmark{2}Faculty of Computer Science, University of New Brunswick, Canada}
}

\maketitle

\begin{abstract}
The recognition of dynamic gestures via discrete myoelectric control is becoming an attractive input modality for event-based human-machine interaction. While recent works suggest that discrete cross-user models are possible (i.e., no calibration data is required from the end user), few studies have examined how to optimize these systems, and how well they generalize to randomly prompted data that simulates real-world use remains largely unknown. Using the 612-user EMG-EPN-612 dataset, this work benchmarks five temporal deep-learning approaches, evaluates the tradeoff between performance and model complexity, and compares hand-crafted and machine-learned feature extraction techniques. Once well configured, the LSTM, BiLSTM, GRU, and Transformer performed similarly (92.8--93.5\%) when tested on 122 previously unseen users, suggesting that the choice of temporal architecture is secondary. These models were then evaluated on a different, more realistic, unsegmented dataset, wherein users were randomly prompted to elicit discrete gestures during a continuous data collection. This dataset was recorded five years later and in a different country from the training data. The cross-user models generalized to this new dataset, with no significant differences between the best configuration of each architecture, even though no feedback was given to users during the elicitation of gestures. These findings suggest that, for discrete zero-shot cross-user myoelectric control, modelling the complete gesture may matter more than the choice of temporal architecture.

\end{abstract}


\section{Introduction}
As human-machine interaction (HMI) continues to expand beyond conventional computing environments, there is a growing need for new input technologies to complement traditional inputs. Leveraging the physiological electromyogram (EMG) signal, the electrical potentials generated by muscles when they are contracted, myoelectric control is a promising option for these emerging use cases \cite{meta_paper}. However, despite its promise and history of success in controlling powered prostheses, the widespread use of this technology for HMI has largely been hindered by the need for regular and personalized end-user calibration \cite{eddy2023framework}. 

Compared to the traditional use case of continuous myoelectric control, whereby decisions are made on small windows of data \cite{kevin_continuous}, discrete control maps dynamic gestures (e.g., a double tap or finger snap) to event-driven commands \cite{eddy2023framework}. Although developing true cross-user models that require no training data from the end user is challenging for continuous control, recent work is beginning to show that the temporal structure of dynamic gestures may help overcome this challenge \cite{eddy2024big, meta_paper, lopez2024cnn}. Moreover, discrete control might even be more appropriate for HMI as many inputs, such as button clicks and swipes, are inherently event-driven \cite{eddy2023framework, lundsberg2025muscle, saponas}.

Although previous work has demonstrated the feasibility of cross-user discrete control without recalibration \cite{eddy2024big, meta_paper, lopez2024cnn}, little work has evaluated techniques and approaches to optimize these systems. This work correspondingly benchmarks five temporal deep learning architectures and compares handcrafted feature extraction to machine-learned CNN-based approaches on the 612-user EMG-EPN-612 dataset \cite{emgepn612}. Additionally, to approximate real-world use, a novel dataset with 14 participants was used, where users were randomly prompted to elicit discrete gestures during a continuous data collection. This required that these gestures be detected and segmented from the continuous input stream, which is often ignored in offline studies. The results suggest that even under these more realistic settings, well-configured models of each architecture generalize to data from unseen users. These results provide valuable insights for the future development of discrete cross-user myoelectric control systems and further highlight the potential viability of this promising technology.

\section{Methods}

\subsection{Datasets}
This work leveraged two datasets that were recorded nearly five years apart: (1) EMG-EPN-612 \cite{emgepn612} and (2) MyoDisCo \cite{eddy2024understanding}. Each dataset was recorded with the Myo Armband, an 8-channel surface-EMG device with a sampling rate of 200 Hz \cite{myo}. The armband was placed in approximately the same location on each subject, with the LED on the extensor and the USB port facing down toward the hand. The \textit{six} common gestures used here were: (1) hand open, (2) hand close, (3) wrist flexion, (4) wrist extension, (5) double tap, and (6) rest.

\subsubsection{EMG-EPN-612}
Recorded by Benalcazar \textit{et al.}, this is one of the largest publicly available EMG datasets by number of participants \cite{emgepn612, benalcazar2026dataset}. During the data collection, users were prompted to elicit each discrete gesture at some point during a five-second window (this included the ramp on and off of the gesture). Indices indicating the start and end of each gesture are included in the dataset's metadata. Labelled data are available for 50 repetitions of each gesture from 306 users and 25 repetitions from the other 306 (a total of 137,700 templates, including rest). A more detailed description of this dataset can be found in  \cite{eddy2024big} and \cite{vasconez2023comparison}.

\subsubsection{MyoDisCo}
\begin{figure}[h!]
    \centering
    \includegraphics[width=\linewidth]{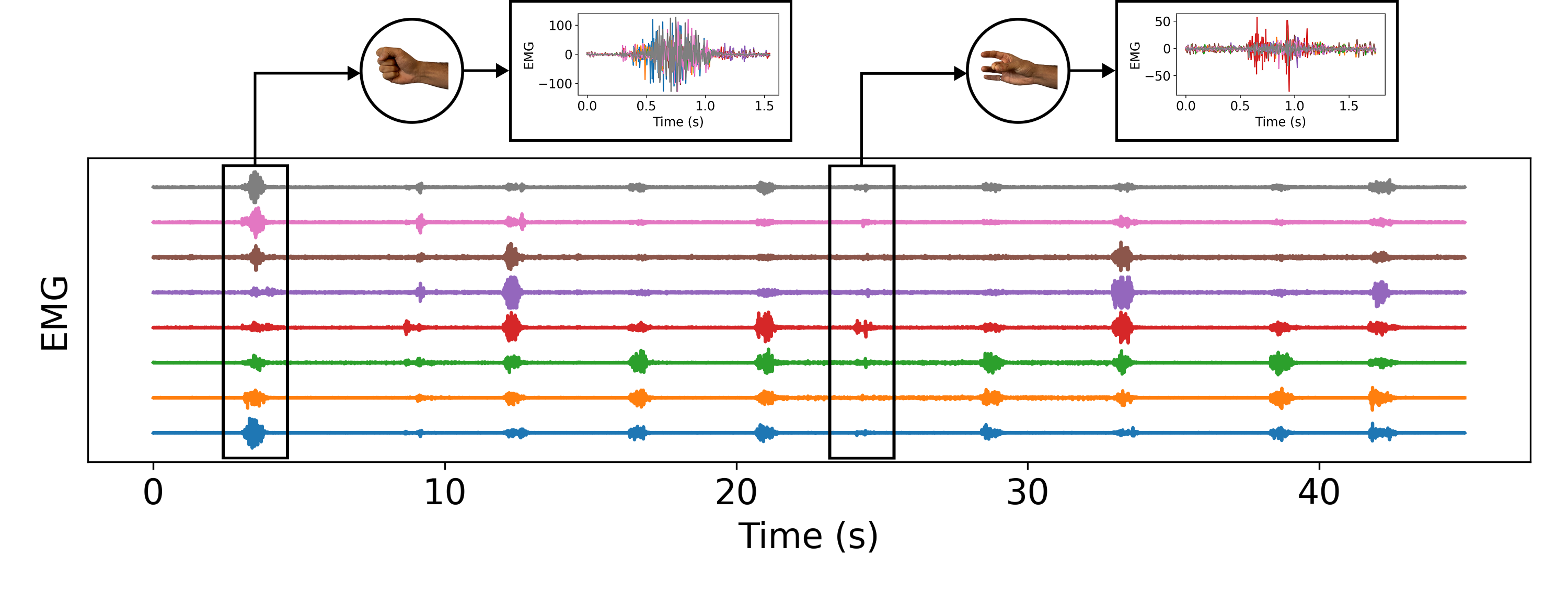}
    \caption{A 45-second example from one subject showing the raw EMG data recorded during the continuous data collection.}
    \label{fig:collection}
\end{figure}

The MyoDisCo dataset was initially recorded to evaluate the impact of confounding factors for discrete myoelectric control \cite{eddy2024understanding}. A portion of this dataset that had not previously been analyzed is the continuous gesture collection (see Figure \ref{fig:collection}). During this collection, participants were randomly prompted every 3-5 seconds (by a pop-up on a computer screen) to elicit one of the discrete gestures. No feedback was given to participants during this session. This collection aimed to provide a more real-world benchmark to evaluate the performance of discrete models when the active gesture must be segmented from the continuous EMG stream. This data collection was approved by the University of New Brunswick Research Ethics Board (REB 2023-140).

\subsection{Classification}
The deep learning architecture used for classifying the discrete gestures was inspired by the two closest previous works \cite{eddy2024big, meta_paper}. The data for each discrete gesture in the dataset were first processed and windowed (size=50 ms and increment=25 ms). For the handcrafted feature case, the wavelet energy (WENG) feature was extracted from each window \cite{weng}. For the machine-learned feature extraction case, three 1D convolutional layers (kernel size of 3 and stride of 1) were applied to the raw windowed data with output channel sizes of 16, 32, and 64. Each layer maintained constant padding and had batch normalization, max pooling, a ReLU activation function, and dropout of 0.2. 
Each template (padded with trailing zeros to ensure constant template sizes in the dataset) was then passed through a network of 1 to 7 temporal layers each with a hidden size of 128 and dropout of 0.2. Only the final output encodings from the last window of the gesture were forwarded to the classification network and used to update the network. The classification network was composed of three linear layers of size 128, 64, and 32 before being projected to a final layer of 6 neurons representing the 6 possible classes. ReLU activation functions were added between each linear layer. An Adam optimizer \cite{kingma2014adam} and the cross-entropy loss function were used in all cases. Additionally, gradient norm clipping was employed to mitigate the issue of exploding gradients during training. Each model was trained for 50 epochs with an initial learning rate of 0.001 and a batch size of 1000. The labelled EMG-EPN-612 data were split by user, with 490 users (80\%) used for training and 122 (20\%) for testing. No additional tuning or training was done when evaluating the models with the MyoDisCo dataset. All code was developed in Python and leveraged \texttt{PyTorch} \cite{paszke1912pytorch} for deep learning, and \texttt{LibEMG} \cite{eddy2023libemg} for EMG-specific data processing.

\subsubsection{Algorithms}
This work evaluated five temporal architectures. The LSTM, bi-directional LSTM (BiLSTM), and GRU all had identical structures as highlighted above. For the RNN, however, the nonlinearity added between each layer was modified from Tanh to ReLU to account for vanishing gradients that were observed during initial training attempts. The transformer was designed with a $d\_model$ of 128 to have a similar hidden representation size as the hidden layers of the other models. The $dim\_feedforward$ was set to 512, the number of attention heads to eight, and positional embeddings were learned during training.

\subsection{Gesture Segmentation}
As participants returned to rest between gestures in the continuous data collection, an energy-based thresholding approach could be used to extract potential gesture templates from the continuous stream of EMG data \cite{virtual_trackpad, chen2013pattern}. 
The Teager-Kaiser energy was taken across each channel of EMG data and then averaged and smoothed using a window size of 150 ms with an increment of 25 ms \cite{solnik2008teager}. The first step in the gesture extraction process involved waiting until the energy fell below a desired threshold $t_{ready}$. Next, a gesture was assumed to have started when the energy surpassed an onset threshold $t_{onset}$. The gesture extractor then waited until the energy reached its maximum amplitude, with a buffer (i.e., it waited up to 100 ms without an increase before stopping). Finally, the gesture detector waited until the energy reached a local minimum and the extracted gesture was passed through the classifier. Based on pilot studies, $t_{ready}$ and $t_{onset}$ were set to 75 and 150, respectively. 

\subsection{Statistical Analysis}
Using a Shapiro-Wilk test, all data were first checked to see if they followed a normal distribution. As the data were not normally distributed, Friedman tests were used to check for significant differences between the factors of interest and, post-hoc, paired Nemenyi tests were used to check for significant differences between specific pairs. A non-parametric Wilcoxon test was applied for any comparisons where the number of groups was less than three. 

\section{Results}

\subsection{Offline Results}
\begin{figure*}[t!]
    \centering
    \includegraphics[width=\linewidth]{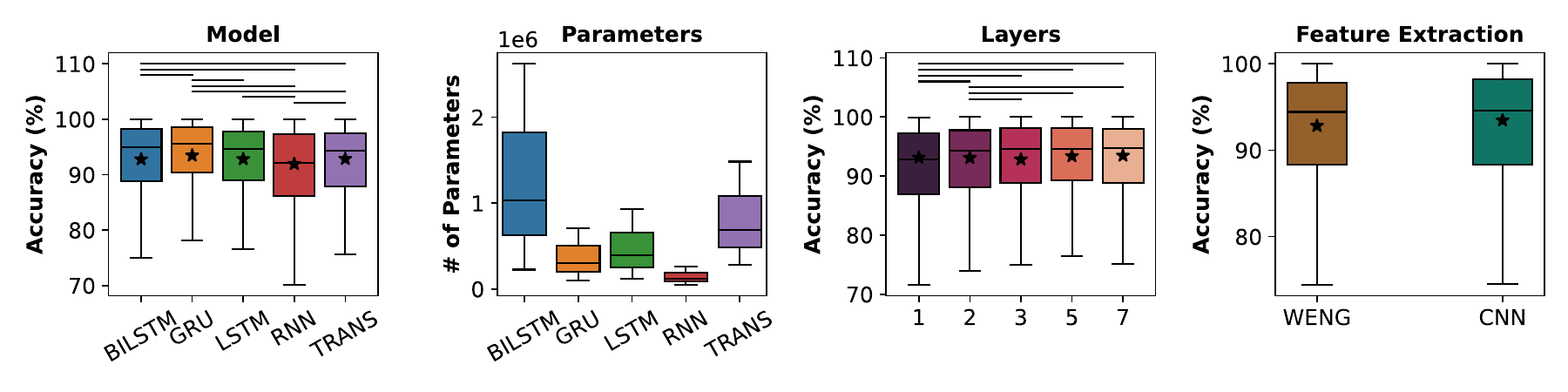}
    \caption{The offline results extracted from the EMG-EPN dataset. The number of parameters plot shows the distribution across the different configurations for each model. All other plots show the distribution for each test subject averaged across the various configurations. The $\star$ indicates the average performance across test subjects for the best configuration for each condition. }
    \label{fig:offline}
\end{figure*}

The results (see Figure \ref{fig:offline}) focus on the dependent variable \textit{classification accuracy} and three independent variables: (1) \model, (2) \features, and (3) \layers. 

\subsubsection{Model}
When averaging across {\layers} and {\features}, a significant effect of {\model} was found \statinfo{(\chisquare{4}{248}{2.2 \times 10^{-16}})}. Pairwise comparisons highlighted significant differences (of \statinfo{p$<$0.001 } unless specified) between the following: BiLSTM \statinfo{(92.1\%)} and GRU \statinfo{(92.9\%)}, BiLSTM and RNN  \statinfo{(89.8\%)}, BiLSTM and Transformer \statinfo{(91.6\%)} \statinfo{(p=0.023)}, GRU and LSTM \statinfo{(91.9\%)}, GRU and RNN, GRU and Transformer, LSTM and RNN, and finally, RNN and Transformer. 

When taking the best configuration for each model, a significant effect of {\model} was found \statinfo{(\chisquare{4}{23}{0.0001})}. Significant differences were found between RNN (layers=5, cnn=False) \statinfo{(91.9\%)} and LSTM (layers=3, cnn=False) \statinfo{(92.8\%)} \statinfo{(p=0.009)} and RNN and GRU (layers=7, cnn=True) \statinfo{(93.5\%)} \statinfo{(p=0.0003)}. No significant differences were found for the BiLSTM (layers=2, cnn=True) \statinfo{(92.8\%)} or Transformer (layers=7, cnn=True) \statinfo{(92.8\%)}.

\subsubsection{Parameters}
The models in ascending order of average number of parameters were the RNN \statinfo{(140916)}, GRU \statinfo{(360257)}, LSTM \statinfo{(469927)}, Transformer \statinfo{(804033)}, and BiLSTM \statinfo{(1265780)}. 

\subsubsection{Layers}
A significant effect of {\layers} was found when averaging across {\features} and {\model} \statinfo{(\chisquare{4}{215}{2.2 \times 10^{-16}})}. Pairwise comparisons highlighted significant differences (\statinfo{p$<$0.001} unless noted) between the following: 1 \statinfo{(90.6\%)} and 2 \statinfo{(91.6\%)}, 1 and 3 \statinfo{(92.0\%)}, 1 and 5 \statinfo{(92.1\%)}, 1 and 7 \statinfo{(92.0\%)}, 2 and 3, 2 and 5, and finally, 2 and 7 \statinfo{(p=0.009)}. No significant effect of {\layers} was found when comparing the best configuration from the different number of layers \statinfo{(\chisquare{4}{4}{0.34})}.

\subsubsection{Feature Extraction}
When averaging across {\model} and {\layers}, no significant differences \statinfo{(p=0.052)} were found between the handcrafted (WENG) \statinfo{(91.4\%)} and machine-learned (CNN) \statinfo{(91.9\%)} features. Additionally, no significant differences \statinfo{(p=0.29)} were found when comparing the best configuration with handcrafted (model=GRU, layers=5) \statinfo{(92.8\%)} and machine learned (model=GRU, layers=7) \statinfo{(93.5\%)} features.

\subsection{Continuous Stream Evaluation}
\begin{figure}[h!]
    \centering
    \includegraphics[width=0.8\linewidth]{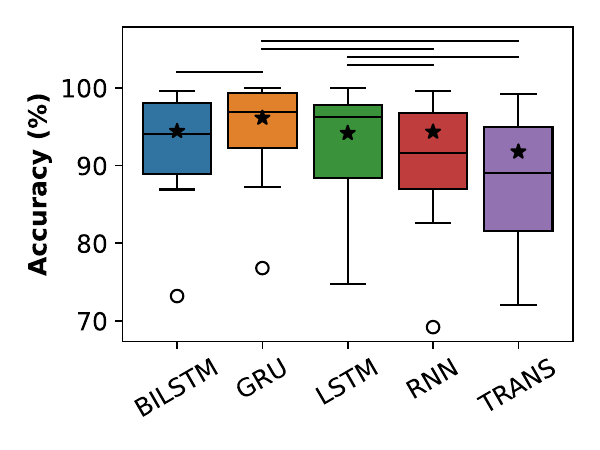}
    \caption{The performance of the cross-user models on the MyoDisCo dataset for each subject averaged across the different configurations. The $\star$ indicates the average performance across subjects for each model's best configuration.}
    \label{fig:online}
\end{figure}

The results from the MyoDisCo dataset are shown in Figure \ref{fig:online}. Using the energy-based gesture extraction strategy, false positive and negative rates (for detecting gestures) were found to be 1.7\% and 0.3\% with an average delay from prompt to prediction (including the elicitation of the gesture) of 700 ms. When averaging across {\layers} and {\features} a significant effect of {\model} was found \statinfo{(\chisquare{4}{32}{1.6 \times 10^{-6}})}. Pairwise comparisons highlighted significant differences between the following: BiLSTM \statinfo{(92.7\%)} and GRU \statinfo{(94.4\%)} \statinfo{(p=0.028)}, GRU and RNN \statinfo{(90.4\%)} \statinfo{(p$<$0.001)}, GRU and Transformer \statinfo{(88.5\%)} \statinfo{(p$<$0.001)}, LSTM \statinfo{(92.8\%)} and RNN \statinfo{(p=0.028)}, and finally, LSTM and Transformer \statinfo{(p=0.009)}. No significant differences were found \statinfo{(\chisquare{4}{3}{0.5})} when comparing the best configuration from each model: BiLSTM (layers=3, cnn=True) \statinfo{(94.4\%)}, GRU (layers=2, cnn=True) \statinfo{(96.1\%)}, LSTM (layers=2, cnn=True) \statinfo{(94.2\%)}, RNN (layers=5, cnn=False) \statinfo{(94.3\%)}, and Transformer (layers=3, cnn=True) \statinfo{(91.8\%)}.

\section{Discussion}
This paper evaluated five temporal deep learning architectures for discrete myoelectric control. Averaged across configurations, the GRU performed best on both datasets, but the ranking of the remaining architectures differed between datasets. Subtle differences in the generalization of some models (e.g., BiLSTM) to the MyoDisCo dataset may be due to differences in how gestures were segmented (it is unknown how segmentation was achieved in the EPN dataset). This suggests that minor variations in the framing of input templates may affect model robustness. While initial differences were significant when averaging across configurations, these results were less pronounced when comparing the optimized configuration from each model. This suggests that all of the proposed architectures are viable if configured correctly. This is consistent with Barona L\'opez \textit{et al.}, who found on the EMG-EPN-612 dataset that adding an LSTM to a CNN improved classification accuracy by only 1.4 percentage points, at the cost of 53 times more learnable parameters \cite{lopez2024cnn}. Because no separate validation set was used, the best configuration for each dataset was identified using its own test users, so these figures should be interpreted as a comparison between architectures rather than as an estimate of the accuracy of a deployed model.

This evaluation had several limitations. The unsegmented evaluation included only 14 participants, recorded in a single laboratory session with the arm in one posture. Gestures were prompted every 3 to 5 seconds and participants returned to rest between prompts, so the EMG stream contained little other hand or arm activity. The 1.7\% false positive rate therefore reflects gesture detection in a relatively quiet stream, not during activities of daily living, where false activations are far more common. The confounding factors known to affect discrete control, such as limb position and cross-day use, were also not introduced. The continuous recordings were processed offline, so participants received no feedback and could not adapt to the model, as they would during online use. The segmentation thresholds were set from pilot data, and because the method used to segment the EMG-EPN-612 gestures is unknown, templates in the two datasets may have been framed differently. Finally, the fifth gesture is labelled ``pinch'' in EMG-EPN-612 and ``double tap'' in MyoDisCo, so the models were evaluated on a gesture that was named, and possibly performed, differently from the one they were trained on.

The BiLSTM and Transformer architectures required substantially more parameters than the other models. As these control systems will likely be deployed on embedded systems, this becomes an important design consideration. While increasing the layers improved performance on the EPN dataset, the best models on MyoDisCo were not necessarily the largest, suggesting that the additional parameters did not lead to improved generalizability. While the CNN-based feature extraction did not significantly improve performance, the highest-performing models on the MyoDisCo dataset leveraged the CNN architecture, suggesting that learned features may generalize well across datasets. Future work could likely optimize these systems further for improved performance. Because well-configured architectures performed similarly and larger models did not generalize better, the choice of temporal architecture for discrete cross-user control can be guided by model size and computational cost, which favours smaller recurrent models such as the GRU for embedded wearable devices.

\bibliographystyle{ieeetr}
\bibliography{bib}

\vfill

\end{document}